\documentclass[twocolumn,superscriptaddress,showpacs]{revtex4}

\usepackage{graphicx}
\usepackage{amsmath}
\usepackage{amssymb}

\DeclareMathSymbol{\ga}{3}{AMSa}{38}

\usepackage{bm}

\usepackage{url}
\usepackage{hyperref}

\newcommand{\Mpch}{$h^{-1}$~Mpc}
\newcommand{\MpcCubeh}{$h^{-3}$~Mpc$^3$}
\newcommand{\kms}{km s$^{-1}$}
\newcommand{\kmsMpc}{km s$^{-1}$ Mpc}

\begin{document}

\title{Lagrangian reconstruction of cosmic velocity fields}

\author{Guilhem Lavaux}
\email{lavaux@iap.fr}
\homepage{http://www.iap.fr/users/lavaux/}
\affiliation{Institut d'Astrophysique de Paris, 98bis Bd Arago, 75015 PARIS, FRANCE}
\affiliation{Universit\'e Paris 6}
\affiliation{Universit\'e Paris 11}

\begin{abstract}
 We discuss a Lagrangian reconstruction method of the
  velocity field from galaxy redshift catalog that takes its root in the Euler equation. This results
  in a ``functional'' of the velocity field which must be
  minimized. This is helped by an algorithm solving the minimization of
  cost-flow problems. The results obtained by applying this method to
  cosmological problems are shown and boundary effects happening in
  real observational cases are then discussed. Finally, a statistical model of the errors made by the reconstruction method is proposed.
\end{abstract}


\maketitle

\newcommand{\LCDM}{$\Lambda$CDM}

\section{\label{intro}Introduction}

Cosmologists are highly interested in studying galaxy
peculiar velocities. Indeed, their study is a direct way to measure
the dynamical state of a system and would thus permit to better
understand dark matter distribution in our local Universe. The main
difficulty is that measured velocities are only available sparsely and
hence does not provide a good probe of the matter distribution. One must then
devise an algorithm that is able to predict, under fair hypotheses,
galaxy peculiar velocities from their present positions, which are their
sky coordinates and their redshift, and compare the result to the measurement.
Jim Peebles \cite{Peebles89} first tried to do full orbit reconstruction by
evolving the present system back in time. This method proved to be
quite accurate for very small volume and number of objects. However,
whenever one tries to reconstruct orbits of a large number of galaxies, the method
fails because the number of plausible solution is blowing up.
A simplification of this problem is presented:
3D galaxy positions are assumed to be known and a simpler
gravitational dynamic model is going to be assumed. We will also
assume that the dynamics of galaxies is mostly driven by collisionless
dark matter particles. 

This proceeding is organized as follows.
In \S~\ref{sec:recons}, we recall the principal result of the
reconstruction method developed by \cite{Brenier2002} (see also the
companion paper Mohayaee \& Sobolevskii, hereafter MS). The method requests the using
of a special fast algorithm to solve the problem. This algorithm is
presented in \S~\ref{sec:auction}. The method is then applied to a dark
matter distribution obtained from a cosmological simulation and the
reconstructed velocities are checked against the simulated ones
(\S~\ref{sec:apply_cosmo}). Finally, a discussion on problems with bad
boundary conditions, as usually met in observational cosmology, is
quickly discussed in \S~\ref{sec:boundary}.

\section{\label{sec:recons}Velocity reconstruction theory}

The theory of velocity reconstruction in cosmology is detailed by
MS.
We recall here the main results. To reconstruct the
peculiar velocity field one must first compute the displacement field
of dark matter particles by solving a Monge-Amp\`ere
equation [Eq.(16) of MS]. We achieve that by minimizing Eq.(17) of MS
in its simplified form using the ``Auction'' algorithm, with $\sigma$ the pairing map and $\mu$ the mass of each particles of the mesh: 
\begin{equation}
    S_\sigma = \mu \sum\limits_{i=0}^N \left({\bf x}_i - {\bf q}_{\sigma(i)}
    \right)^2\label{eq:mini},
\end{equation}
The minimization is conducted over $\sigma$. 
We recall also the Zel'dovich approximation [Eq.(12) of MS] for the velocity field is taking the following form
\begin{equation}
  {\bf v}({\bf x}_i) = \beta \left({\bf x}_i - {\bf q}_i\right) \label{eq:vel_zeldov}
\end{equation}
where $\beta$ is the linear growth factor, which is well approximated by $\beta \simeq \Omega_\text{m}^{9/5}$ when it is computed at redshift $z=0$.

\section{Minimization algorithm\label{sec:auction}}

Direct minimization of Eq.~\eqref{eq:mini} is a computationally
difficult problem [time complexity $O(N!)$].
Fortunately, there exist better alternatives that have been
developped for solving minimal cost flow problems which can be adapted
to our minimal transportation problem.
In particular, we are going to use the ``Auction'' algorithm developed in
\cite{Bertsekas79}. The time
 complexity of this algorithm is of the order of $O(n^{2.25})$ by direct
 performance measurement, with $n$ the particle density\footnote{This
   number is obtained for a given simulation, and particles randomly
   until the desired average density is obtained. The worst case of
   this algorithm is actually $O(N^3)$, if one makes a dense search on
   purely random data.}. The exact constant
 hidden in $O(n^{2.25})$ depends a
 lot on the difficulty of the assignment problem, which means it is
 catalog dependent.

\begin{figure}
  \begin{center}
    \begin{tabular}{cc}
      \includegraphics[width=.47\hsize]{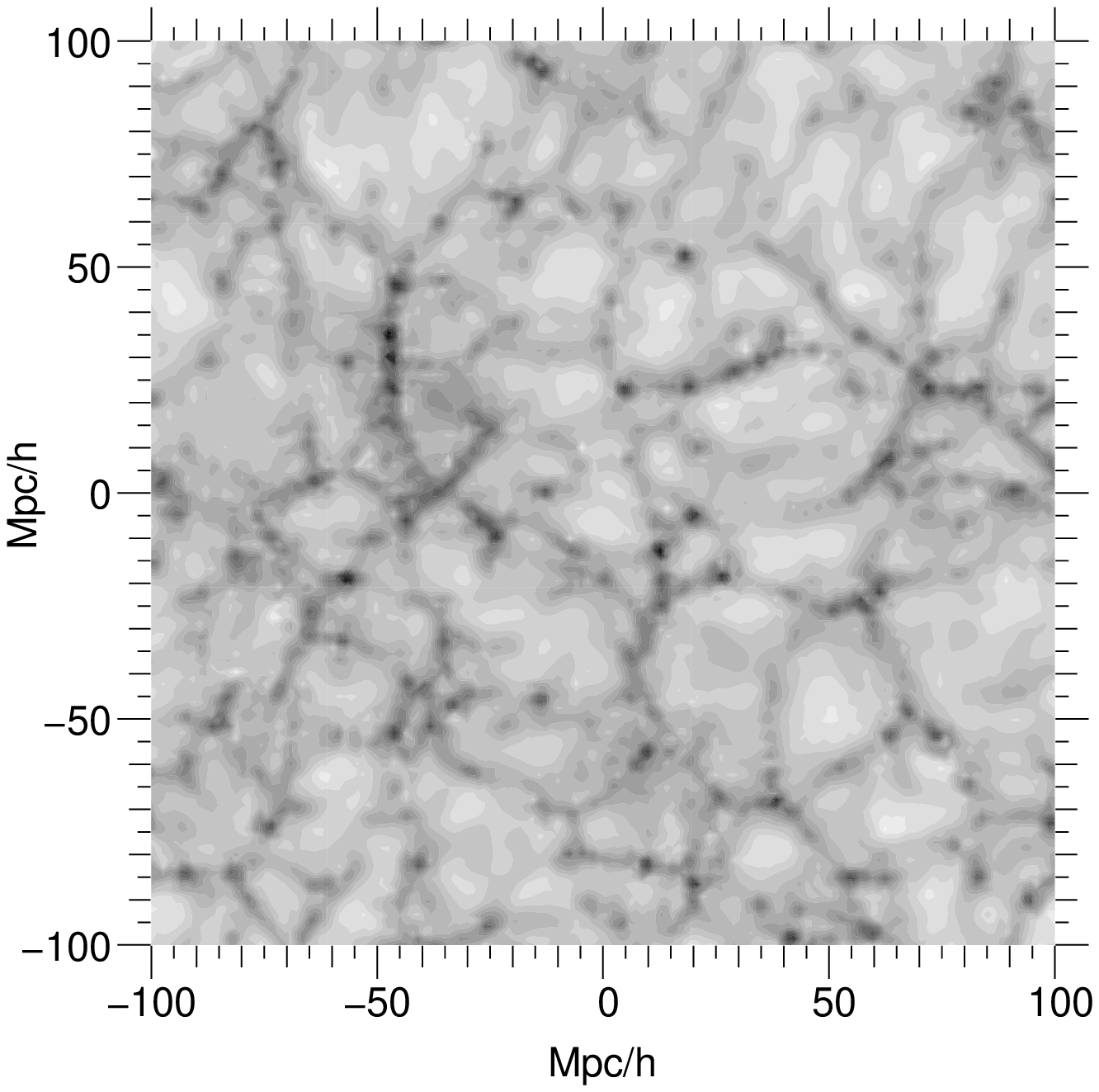} &
      \includegraphics[width=.47\hsize]{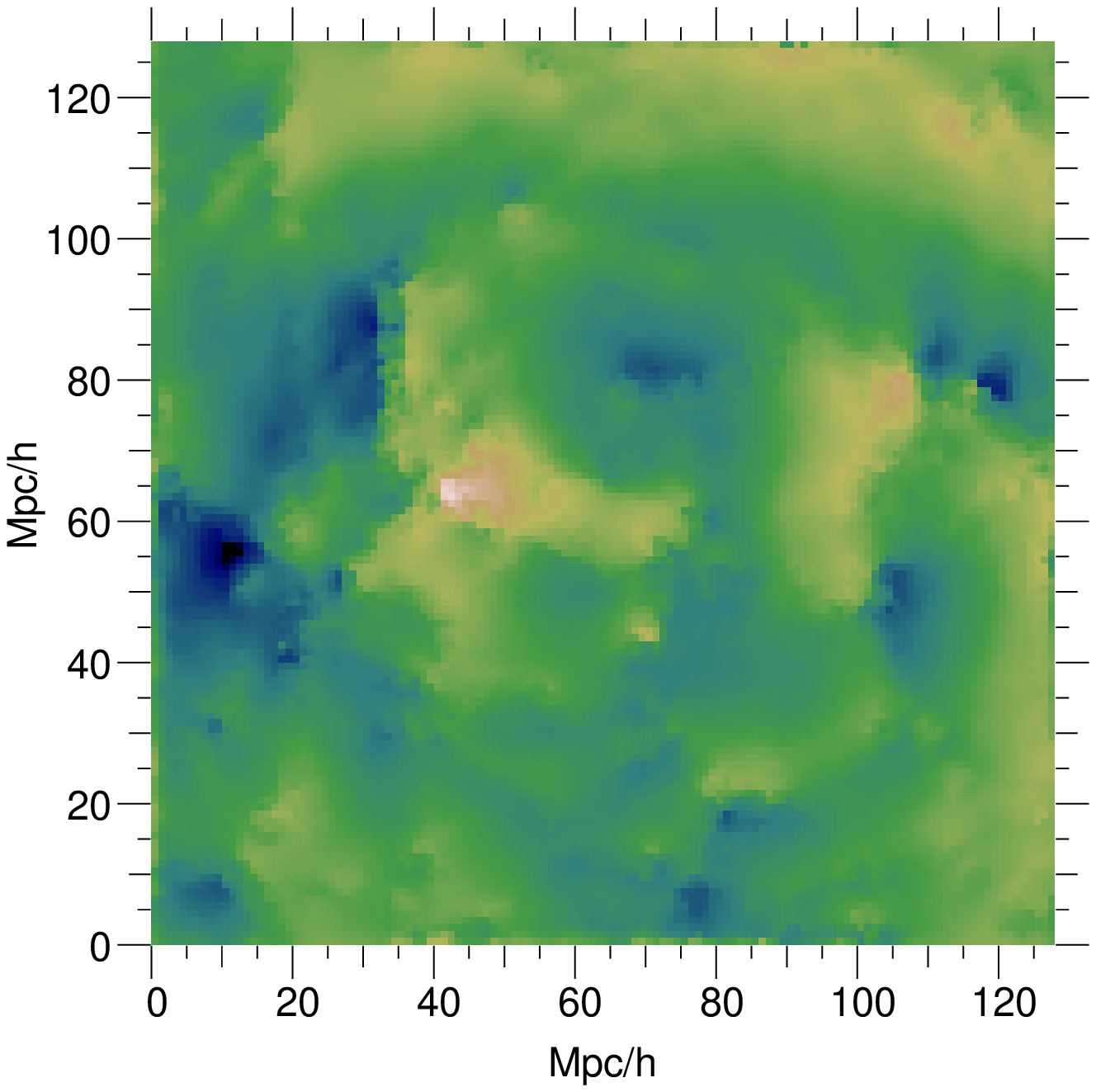} \\
      \includegraphics[width=.47\hsize]{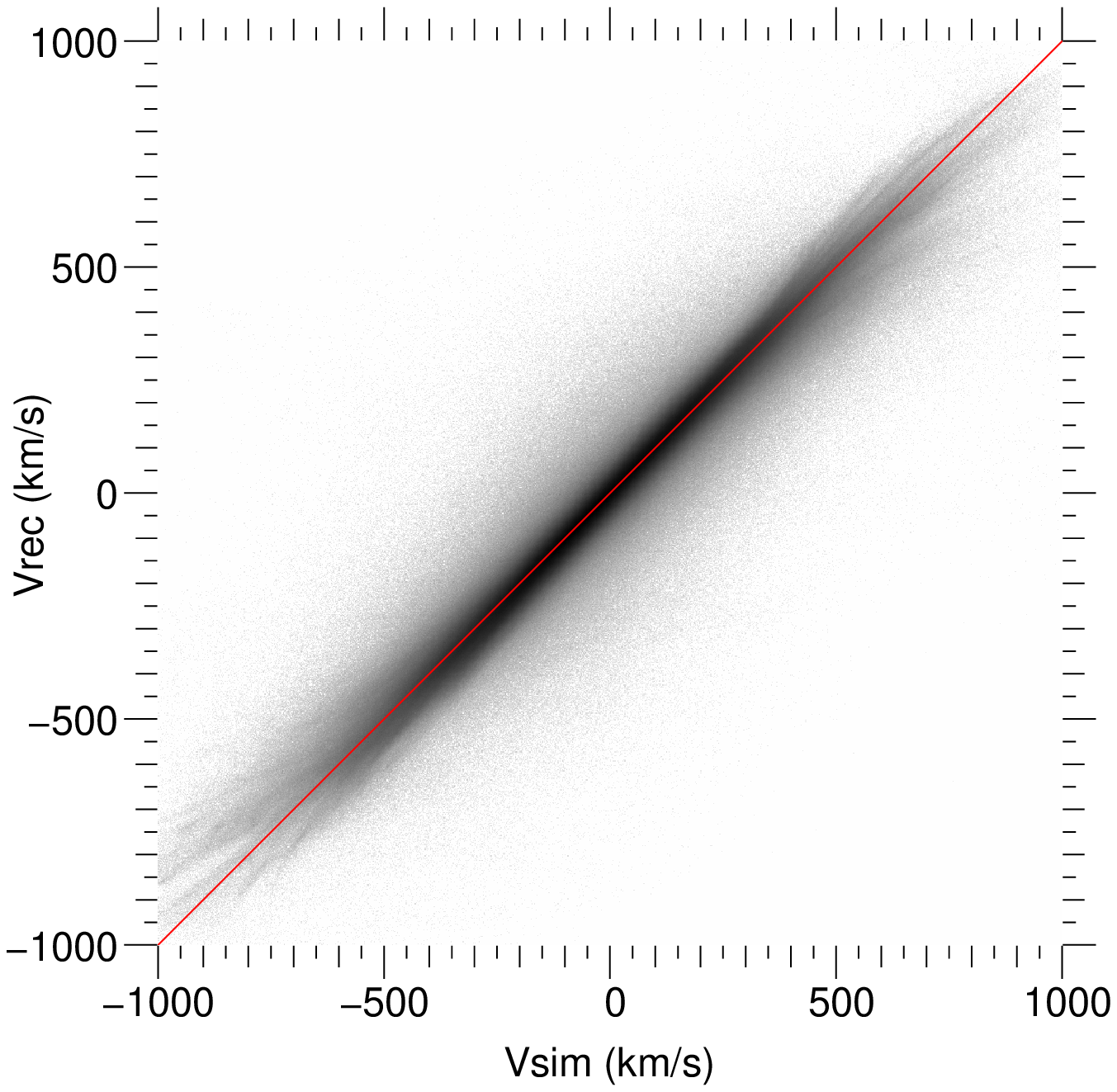}
      & \includegraphics[width=.47\hsize]{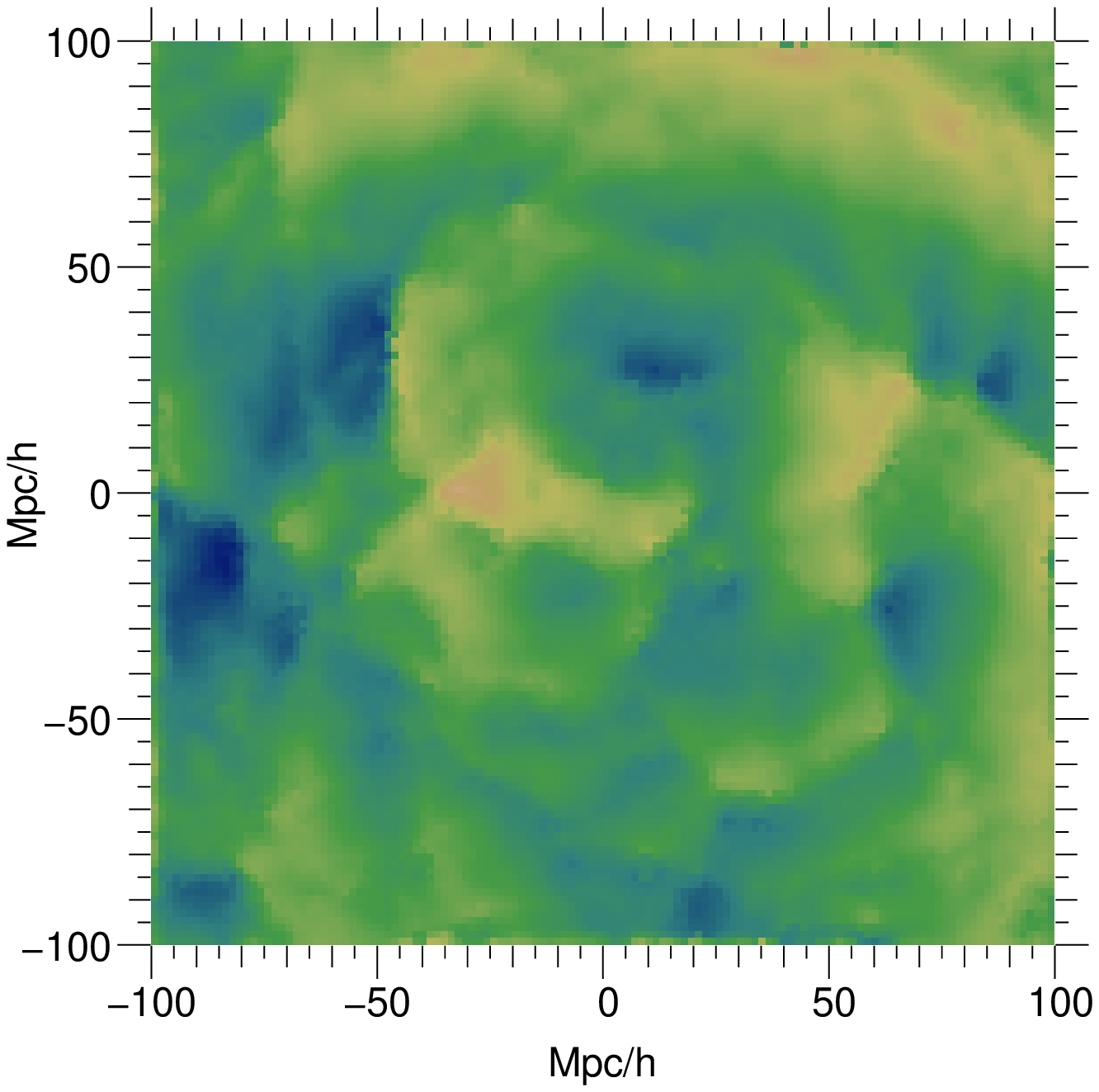} 
    \end{tabular}
  \end{center}
  \caption{\label{fig:simutest} {\it Application to Cosmology} -- {\it Top left}: A slice of the density field of the \LCDM{}
    simulation that is used for the tests (shades of gray indicates
    logarithm of the mass density). {\it Top right}: Adaptively smoothed line-of-sight
    component of the velocity
    field in the same slice.  {\it Bottom right}: MAK reconstructed
    line-of-sight component of the velocity field of
    the same slice. Linear color scale: dark blue=-1000~\kms,
    white=+1000~\kms. 
    {\it Bottom left}: Scatter plot between reconstructed and
    simulated velocities for objects identified in the simulation. 
    Shades of grey show levels of the logarithm of the point density.}
\end{figure}

\subsection{Auction algorithm} 

The algorithm tries to evolve the pairing map $\sigma$ between
${\bf x}_i$ and ${\bf q}_j$ such that when the function is stationary between two
consequent iteration it corresponds to minimizing the given total
association cost. 
Particles locate at different Eulerian positions $x_i$ compete 
against each other for Lagrangian positions $q_j$.
Minimization of the total
association cost $S_\sigma$ is achieved by
studying the dual problem of minimization of association penalities
$p_j$. In \cite{Bertsekas79} it is shown that
\begin{equation}
  \mathop{\text{min}}_\sigma S_\sigma = \mathop{\text{max}}_{
    p_j;j=1,\ldots,n} \left\{ \sum_j p_j + \sum_i r_i \right\},\label{eq:dual}
\end{equation}
with $a_{i,j} = \mu ({\bf x}_i - {\bf q}_j)^2$, the cost of associating
${\bf x}_i$ to ${\bf q}_j$ and $r_i = \text{min}_j (a_{i,j} + p_j)$.
Once the set $\{ p_j \}$ is determined by the above maximization, the
map $\sigma$ is simply given by:
\begin{equation}
  \sigma(i) = \text{arg}\,\mathop{\text{min}}_j \left\{ a_{i,j}
    + p_j \right\}.\label{eq:sigmamin}
\end{equation}
Effectively, $\{ p_j \}$ is computed iteratively by the algorithm. Each iteration
is composed of two parts. During the first one, we obtain a set
of best assignment $\mathcal{A}(j)$ for each particle ${\bf q}_j$ by minimizing all
possible $r_i$. Then, we link ${\bf x}_{i^*_j}$ to ${\bf q}_{j}$ with
$i^*_j$ being the particle having the minimal $r_{i^*_j}$ in the set
$\mathcal{A}(j)$. We also have a reverse mapping for this link that
we write $j^{*}_i$. Finally, the penality $p_j$ is updated such that
\begin{equation}
  p_j \rightarrow \tilde{p}_j = a_{i^*_j, j} + w_{i^*_j} - \epsilon,
\end{equation}
with $\epsilon > 0$ and
\begin{equation}
  w_{i} = \mathop{\text{min}}_{j \neq j^{*}_i} \left(a_{i,j} +
    p_j\right).
\end{equation}
The solution found
is the same as for $\epsilon=0$ provided $\epsilon < \epsilon_0 / N$,
with
\begin{equation}
  \epsilon_0 = \mathop{\text{min}}_{\{i,j\} / a_{i,j} \neq 0} a_{i,j}.
\end{equation}
The time complexity depends quite a lot on the way $\epsilon$ is
scaled down from its initial value to the $\epsilon_0/N$. Numerical
experiments have shown that trying to converge in about 5 iterations
and starting from $\epsilon/\epsilon_0 \simeq N/2$ seems to give a
faster convergence.

\subsection{Implementation}

We developped a C++ multithreaded (shared memory parallelism) and MPI 
version of the ``Auction'' algorithm, it will be available later
as a multi-purpose library for cost-flow problems at the
address \url{http://www.iap.fr/users/lavaux/}.
Besides doing a full minimization over all ${\bf q}_j$ for a given
${\bf x}_i$ (``dense'' mode). It also supports a ``sparse'' mode that
solves a partial minimization problem: for a given ${\bf x}_i$, it
only minimizes over a subset of $\{ {\bf q}_j \}$ such that $||{\bf x}_i - {\bf
  q}_j||_{\infty} < R$, where $R$ is a parameter given at the
initialization to the algorithm. This allows to reduce drastically the
computing time while giving the same result provided that $R$ is not too
small (typically $R=40$~\Mpch{} for a \LCDM{} Universe).
On a Dual-core AMD Athlon64 4800+, the SMP
implementation (dense mode) takes 50 mins to assigning 79,000
particles. It has successfully reconstructed a 128$^3$ dense mesh in a month in sparse mode.
The MPI version of the corresponding algorithm is only performant for
larger number of particles (typically $N \gtrsim 500,000$).
Most of the time is, at the moment, spent at computing
$\mathop{\text{min}}_j \left(a_{i,j} + p_j\right)$ as the cost values are
only kept in a minimalistic cache. Precomputing the costs is also not
feasible because of the excessive amount of memory that would be
needed to store all costs for all $(i,j)$ pairs. We also consider to
implement a general purpose totally asynchronous implementation in the
near future. 

\section{\label{sec:apply_cosmo}Application to cosmology: test on cosmological simulation}

\begin{figure}
  \begin{center}
    \includegraphics[width=.7\hsize]{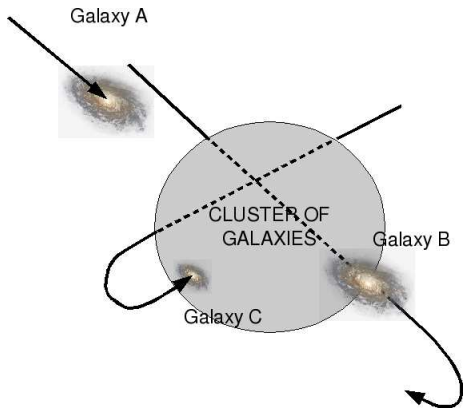}
  \end{center}
  \caption{\label{fig:second_infall} {Cosmology / Multi-streaming regions} -- This figure illustrates the
    different problems that may occur for a halo of dark matter
    particles near a cluster of galaxies. Galaxy A is in the region of
    first infall. The displacement field will be well
    reconstructed. Galaxy B is coming from the same direction as
    Galaxy A but has already gone through the center of the cluster
    and is decelerating. In that case, its displacement is badly
    reconstructed as, most likely, MAK predicts that the matter
    composing Galaxy B is coming from the region opposite to Galaxy
    A's region. Galaxy C is also wrongly reconstructed.}
\end{figure}

To check that the dark matter dynamical model is working, we are
testing it against a 128$^3$ $N$-body sample \citep{moh2005} which was
generated with the public version of the $N$-body code {\sc HYDRA}
\citep{CouHydra95} to simulate collisionless structure formation in a
standard \LCDM{} cosmology. The volume of the simulation is $200^3$\MpcCubeh. The mean matter density is
$\Omega_\text{m}=0.30$ and the cosmological constant is
$\Omega_\Lambda=0.70$. The Hubble constant is $H_0 =
65$~\kmsMpc and the normalization of the density
fluctuations in a sphere of radius 8~\Mpch{} is $\sigma_8=0.99$.

Haloes of dark matter particles are identified using a friend-of-friend
algorithm with a traditional value of the linking parameter 
$l=0.2$ times the mean particle separation. A limit of
5 linked particles is put to bind particles into a halo. The particles
left unbound by this criterion were kept in a set called the
``background field''. All objects are kept in a mock catalogue called
{\it FullMock}. 
We have run a reconstruction on FullMock using a MAK
mesh with $128^3$ elements. Each object of FullMock was given a number
of elements ${\bf x}_i$ equal to the number of particles of the
original simulation which has been bound into this object. We distributed
the ${\bf q}_j$ mesh elements regularly on a cubic grid of the same
physical size as the simulation box. Finally we computed the 
convex mapping $\sigma$ corresponding to the MAK problem with the help
of the algorithm described in \S~\ref{sec:auction}.
The velocities for each
particle were computed using the Zel'dovich approximation
 Eq.~\eqref{eq:vel_zeldov}, using the same
cosmology as the simulation to compute $\dot{D}(t)$.

Fig.~\ref{fig:simutest} summarizes the results obtained using the MAK
method on the reconstructed velocities. The individual object
velocities, in the bottom-left panel, are exceptionally well
reconstructed. Visual inspection of the line-of-sight component of the
velocity field in the two right panels show nearly no discrepancy
except in regions with really high velocities. In these regions, the
dynamics is highly non-linear, which means that the convex hypothesis is
not valid anymore. This problem arises on a typical cosmological scale 
of at most a few Mpc around large clusters. Indeed, in those
regions the fluid description of dark matter particles completely fail
because the mass tracers may have already crossed the center of the
gravitational attractor and are currently falling back to the
center, as illustrated Fig.~\ref{fig:second_infall}. This renders the
displacement field reconstruction dubious in those cases.

\section{\label{sec:boundary} Application to Cosmology: boundary problems}

\begin{figure}
  \begin{center}
    \begin{tabular}{cc}
      \includegraphics[width=.45\hsize]{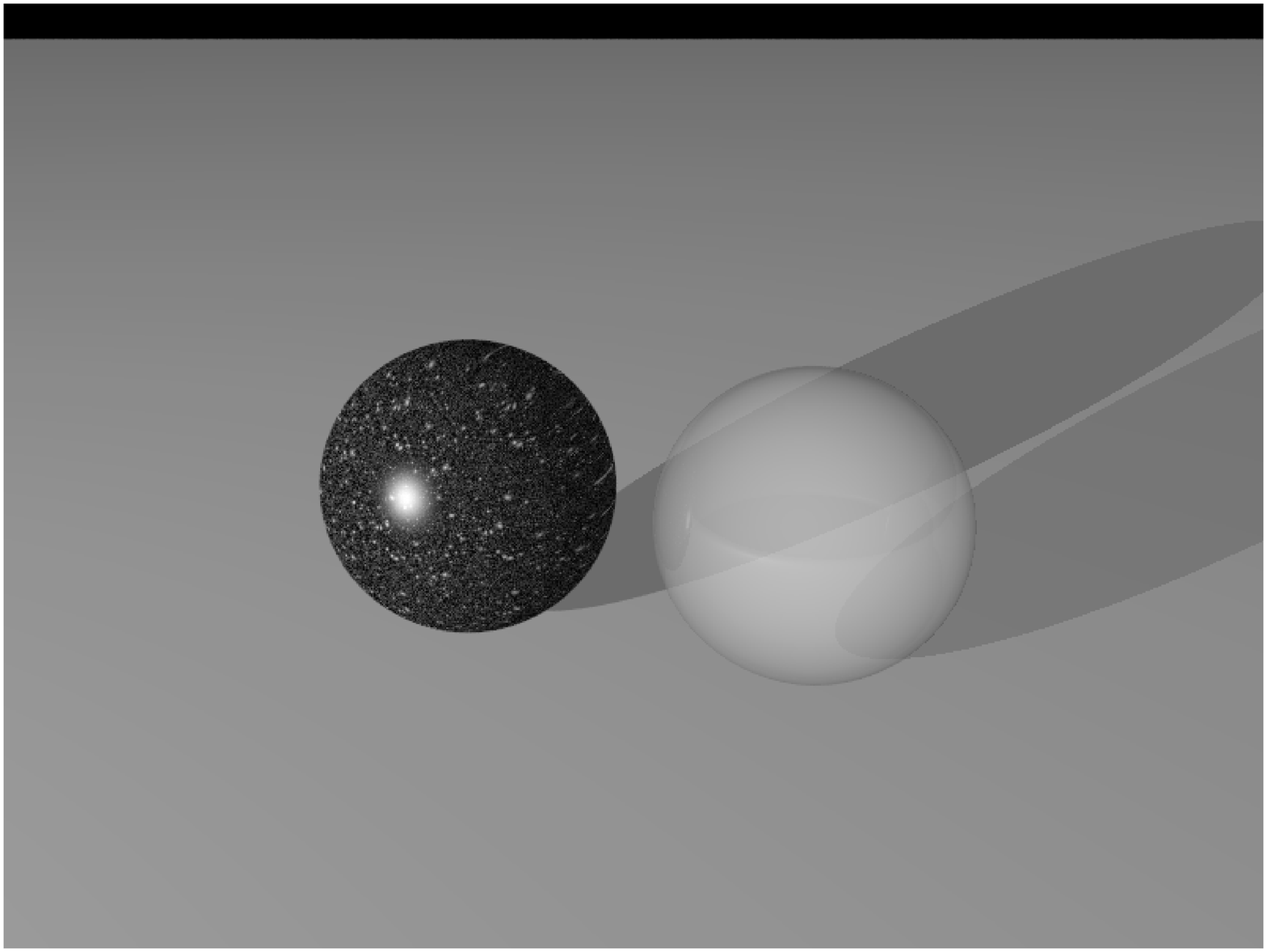} &
      \includegraphics[width=.45\hsize]{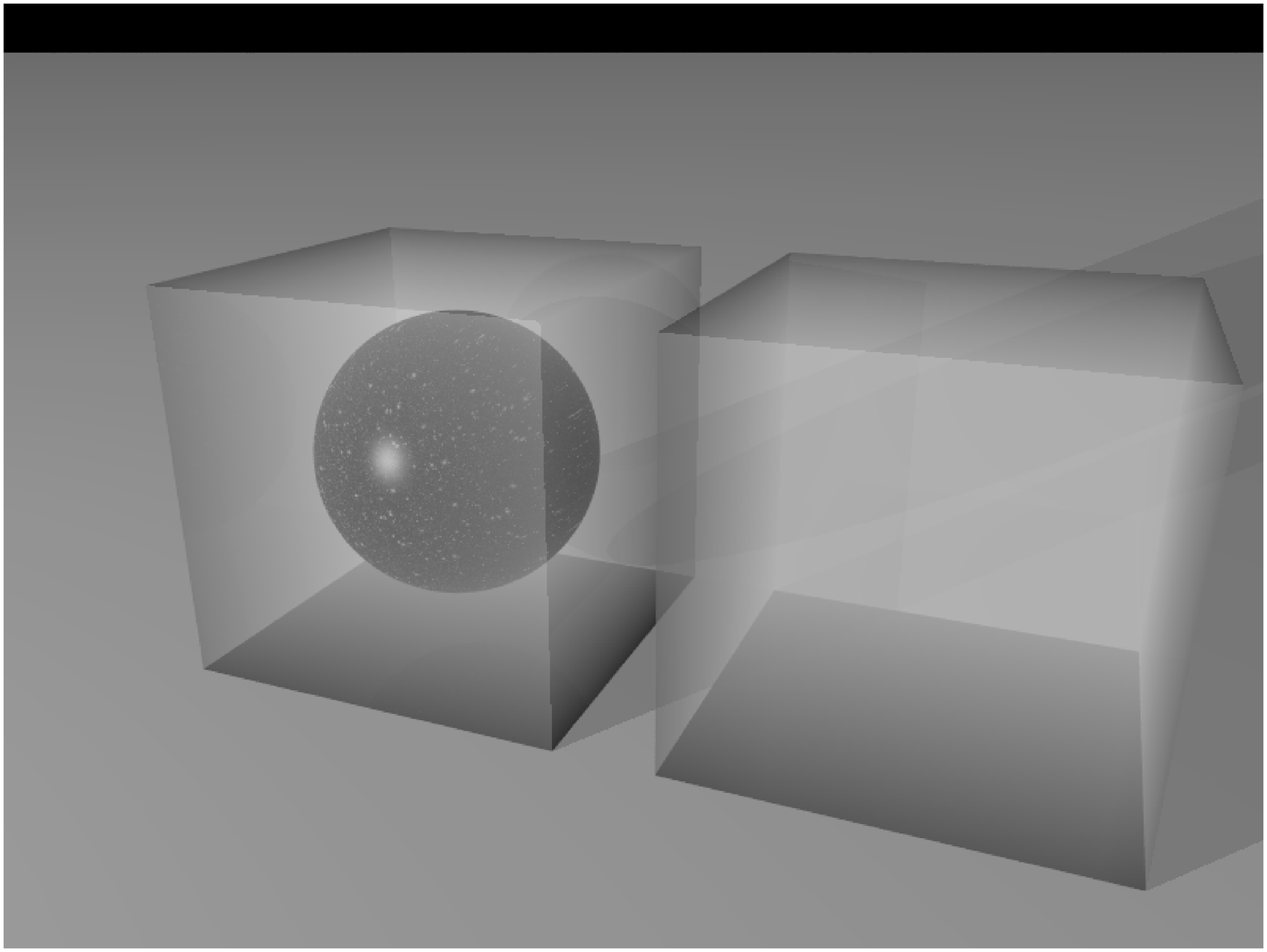}
    \end{tabular}
  \end{center}
  \caption{\label{fig:illustr_bound} {\it Cosmology / Boundary problems} -- {\it Left panel:} Illustration of
    the NaiveDom approach to handle boundary problems while doing a
    reconstruction. The dark starry ball illustrates the current dark matter
    distribution as inferred from galaxy catalogeus. The whitish
    transparent ball is the assumed initial volume for the dark matter
    that has fallen in present structures.
    {\it Right panel:} Same as left panel but this illustrates the
    PaddedDom approach.
  }
\end{figure}

\begin{figure}
  \begin{center}
    \begin{tabular}{cc}
      \includegraphics[width=.47\hsize]{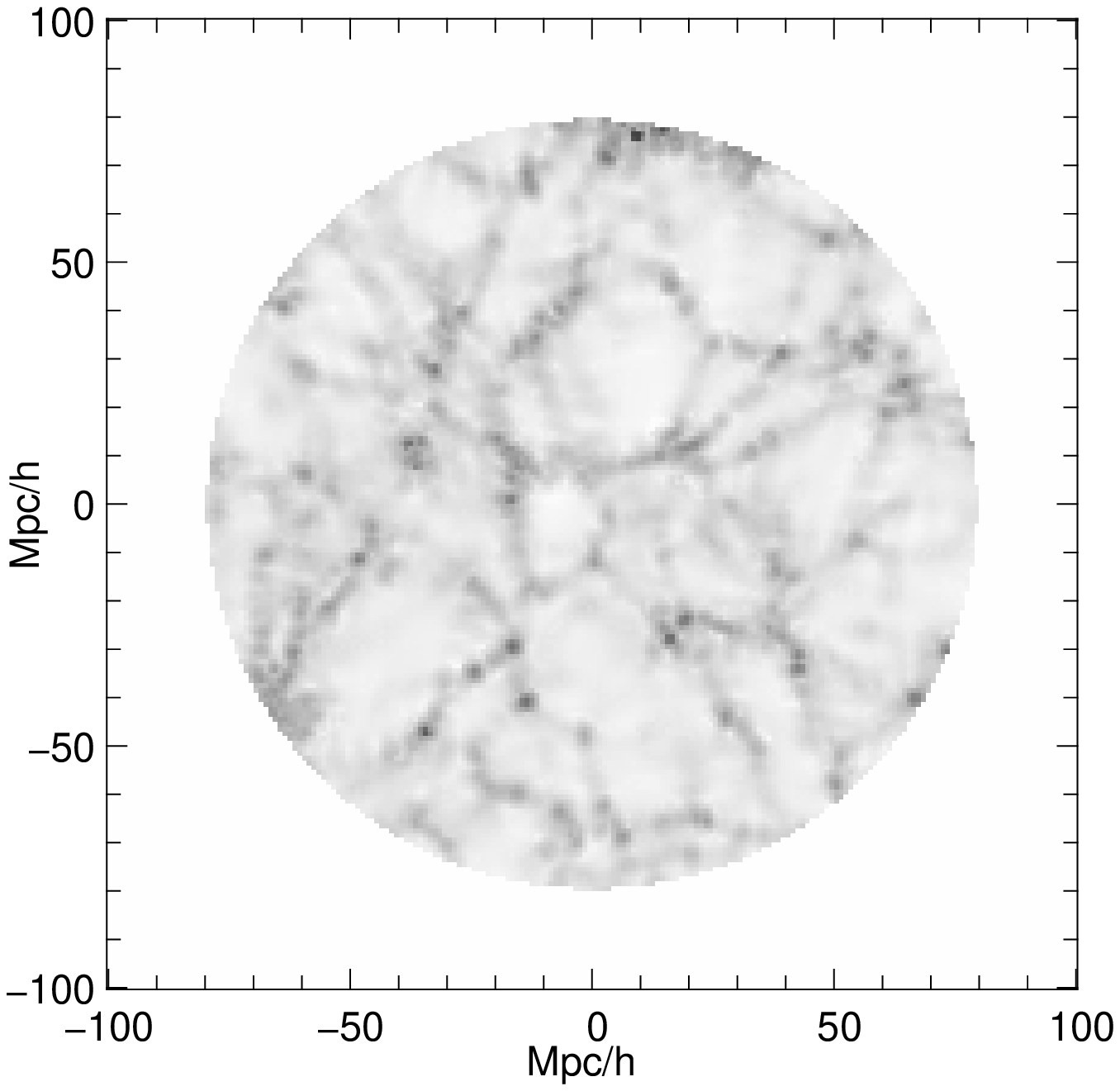} & 
      \includegraphics[width=.47\hsize]{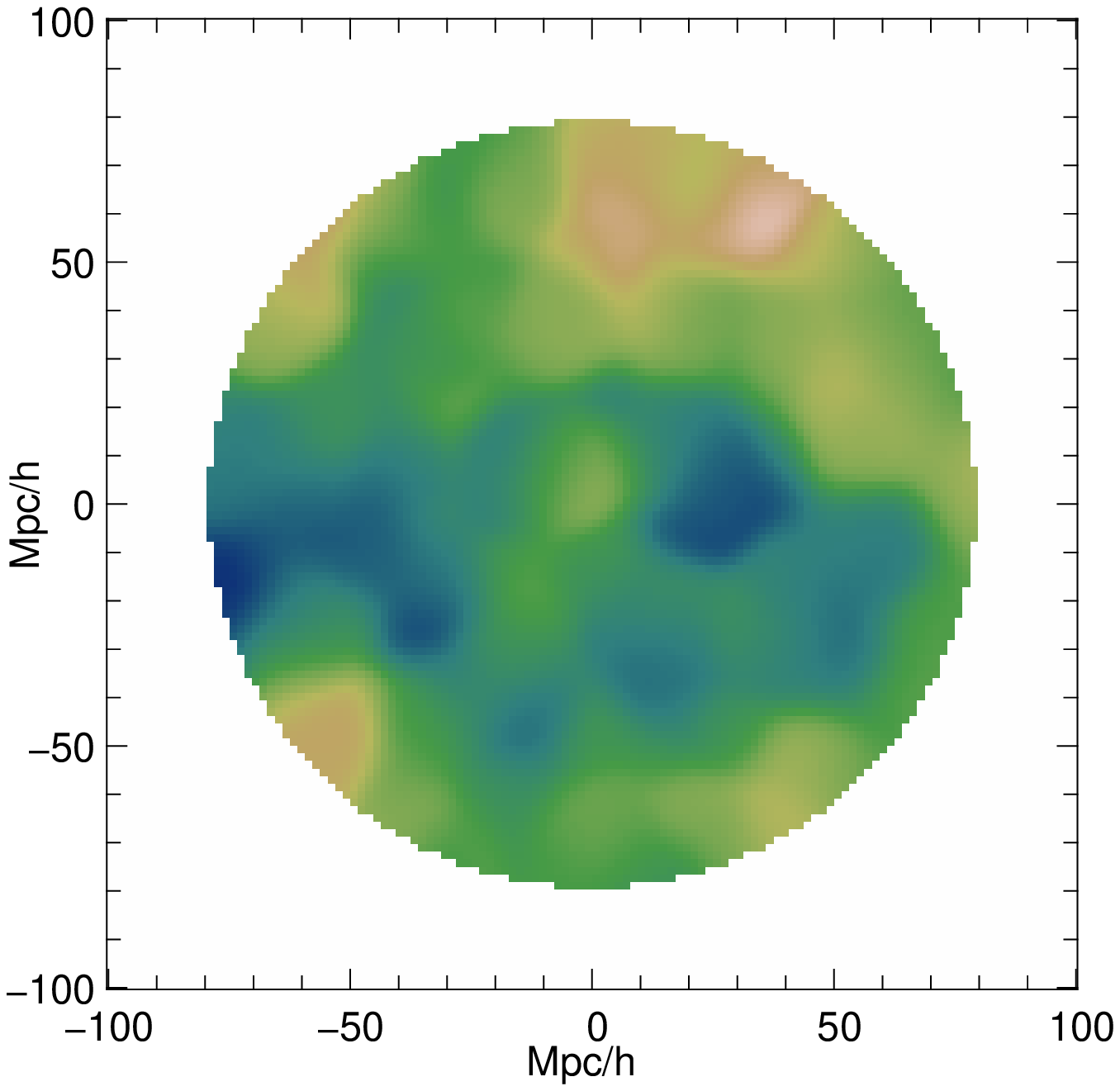} \\
      \includegraphics[width=.47\hsize]{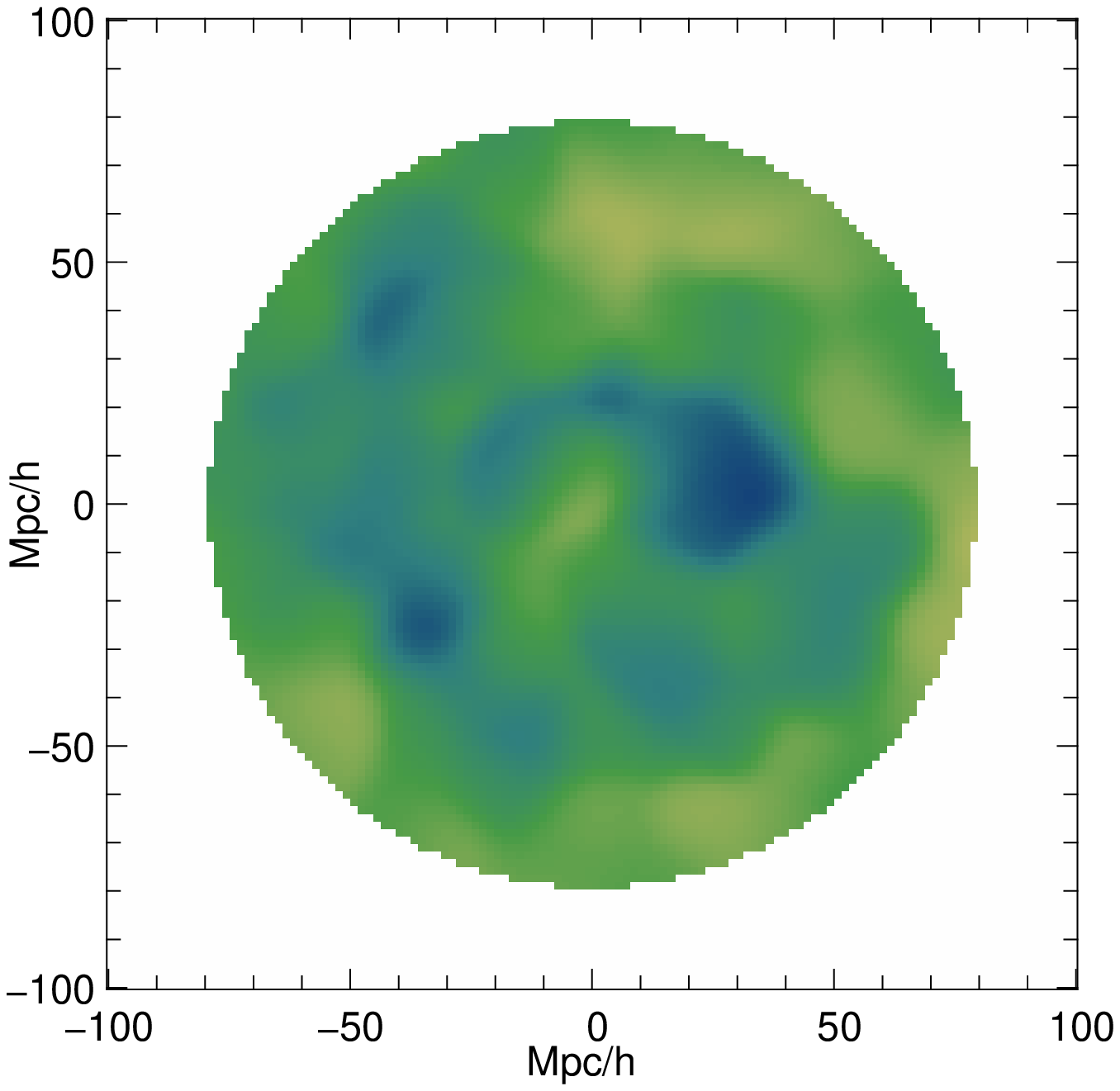}
      &
      \includegraphics[width=.47\hsize]{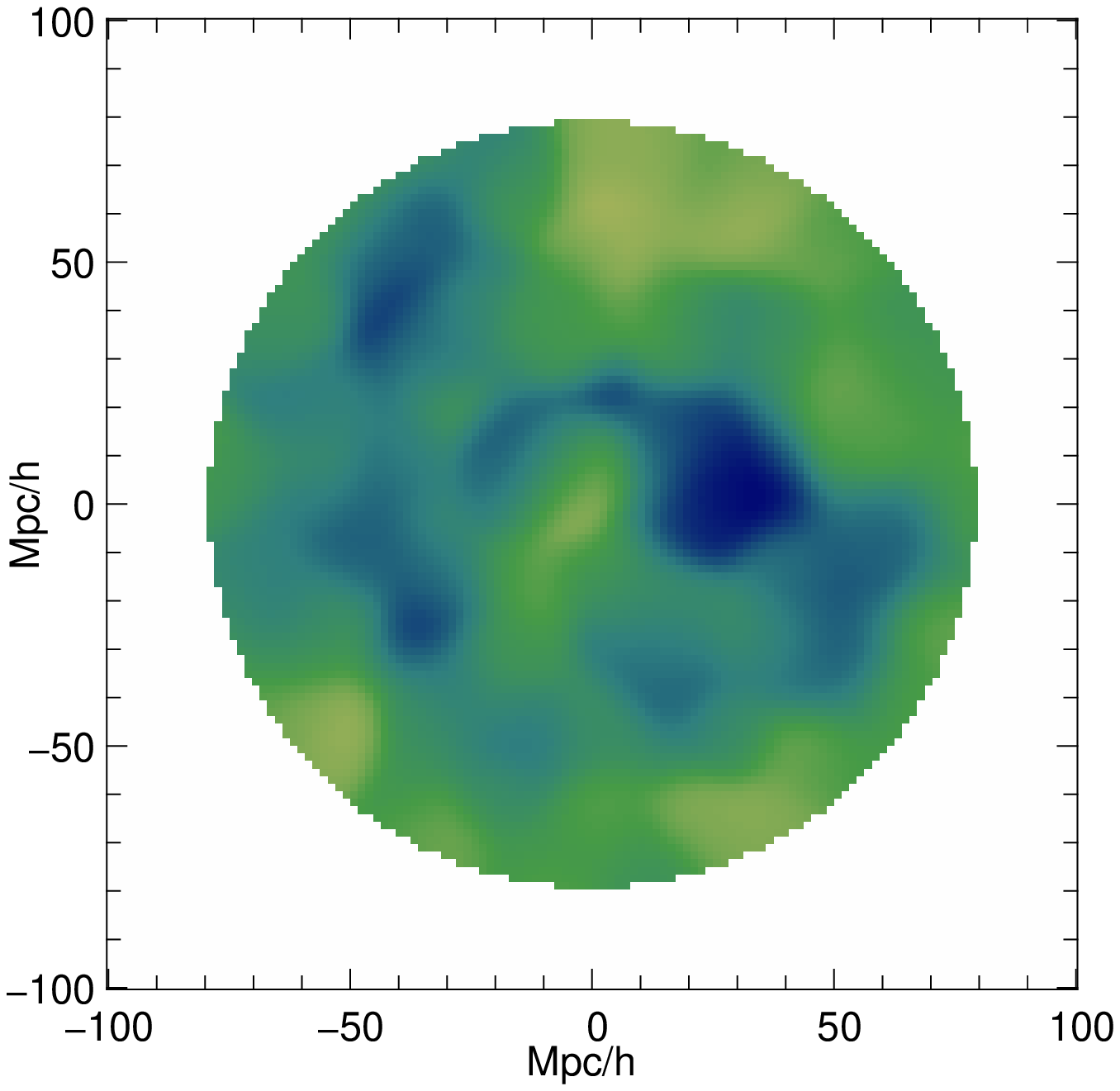}
    \end{tabular}
  \end{center}
  \caption{\label{fig:results_bound} {\it Cosmology / Boundary
      problems} -- Outer boundary problems while doing reconstruction on
    finite volume catalogue. Color scale is the same everywhere (dark
    blue=-1000~km/s, white=+1000~km/s). {\it Top left}: Density
    field of the mock catalogue (log scale). {\it Top right}: Simulated velocity field,
    smoothed with a 5~\Mpch{} Gaussian window. {\it Low left}: PaddedDom velocity
    field, smoothed equally. {\it Low right}: NaiveDom velocity field,
    smoothed equally. }
\end{figure}

One does not necessarily know the Lagrangian domain ${\bf q}$ on which
the MAK reconstruction must be computed. This is the case for real
cosmological observations and one must use some empirical prescription
to attenuate the boundary effects on reconstructed
velocities. This scheme is helped by the overall homogeneity of
the Universe above scales larger than 200~\Mpch. We propose thus to check
two schemes to handle boundary effects:
\begin{itemize}
\item[-] A naive approach would be to assume that the piece of Universe
  considered has not changed its volume sufficiently between initial
  time and the current time. This means that we may assume that if we select
  a ball of matter, in the Universe, centered on us, all the mass that
  is inside this ball is coming from the same homogeneous ball in the Universe
  as it was at decoupling time. We call this approach {\it
    NaiveDom}. It is equivalent to say that tidal field effects are
  totally negligible on the considered scale.
\item[-] An alternative approach is not to make an assumption on the
  exact shape but on the low amount of fluctuation on the boundary.
  Consequently, if one selects the same ball of
  matter in the present Universe, it is fair under this approximation
  to pad the matter distribution using homogeneously distributed
  particles. One may then build the mapping between the ``padded piece
  of Universe'' and an initial completely homogeneous set of
  particles. We call this approach {\it PaddedDom}.
\end{itemize}
These two ways of handling boundary effects are illustrated
Fig.~\ref{fig:illustr_bound} and the results are presented in
Fig.~\ref{fig:results_bound}.

As expected, boundaries are badly reconstructed in
PaddedDom and NaiveDom. However at the center of the spherical cut,
the velocity field seems correctly
reconstructed by visual comparison to the velocity field computed from
the simulation. Looking carefully at the result using NaiveDom
indicates that there is likely a systematic error near the center (the
blue region is darker and more extended than in the two other
figures). This is probably due to stronger boundary effects that are
not correctly attenuated by the NaiveDom scheme (a detailed
  quantitative analysis of boundary artefacts are given in \cite{Lavaux07}).
Empirically, we found that a buffer zone of, at least, about
20~\Mpch{} is needed
to reduce boundary effects with a PaddedDom reconstruction scheme.

\section{\label{sec:stats} Statistical analysis of errors in the reconstruction}

The measurement of the slope between velocities and reconstructed
displacements should give an estimation of $\Omega_\text{m}$. However,
building a reliable estimator of this slope without the statistical
model of errors made both at the observation and the reconstruction
level may produce unaccepable bias. We propose to show how to use models on reconstruction errors to 
make a bayesian analysis of the reconstructed velocities.
We will focus here on errors made during a reconstruction and assume
that the observed peculiar velocities $v$ are equal to their true velocities.
A more detailed discussion can be found in \cite{Lavaux07}.

\begin{figure}[t]
  \includegraphics*[width=.85\hsize]{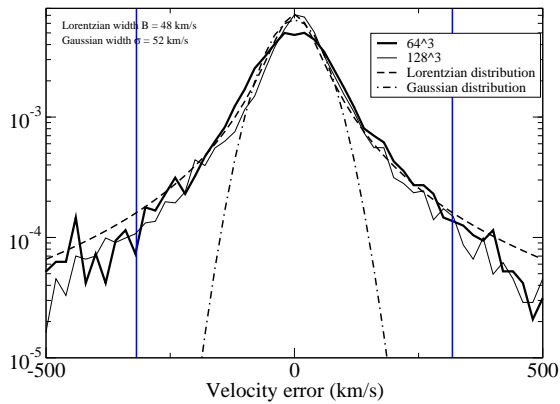}
  \caption{\label{fig:errors} Error in the reconstruction --
    This plot displays the probability distribution of the quantity
    $v_{\text{r,rec}} - v_\text{r,sim}$, where $v_\text{r,rec}$ and
    $v_\text{r,sim}$ are the line-of-sight reconstructed and simulated
    velocities, respectively, after choosing an observer at the center
    of the simulation box. The dashed and dot-dashed curve give the best
    fit of a Gaussian and a Lorentzian distribution, respectively.}
\end{figure}

Using simulations, we have measured the distribution of reconstruction errors, for each object $i$ of a catalog of galaxy, $\{ e_i \}$ defined as
\begin{equation}
  e = v_\text{r} - \beta \psi_\text{r,rec}\;,
\end{equation}
with $\beta=0.51$ for the studied simulation (corresponding to $\Omega_m=0.30$), $v_\text{r}$ the line-of-sight
component of the simulated velocity of the considered, $\psi_\text{r,rec}$ the reconstructed radial displacement.
The result is given in Fig.~\ref{fig:errors}.
We have tried to fit an histogram of the errors $\{ e_i \}$ by both a Gaussian function of
width $B$
\begin{equation}
  f_G(e) \propto \exp\left(-\frac{e^2}{B^2}\right)
\end{equation}
and a Lorentzian function
\begin{equation}
  f_L(e) \propto \frac{1}{1+\frac{e^2}{B^2}}\;.
\end{equation}
We obtained approximately the same width $B$ for the two fits (which is
expected from the second order development of both functions), however
it is striking that $f_L$ is a much better approximation than $f_G$ to
the observed error distribution.

We equate the probability of getting an error $e$ on the true
velocity $v_\text{r}$ for an object of the catalog to $f_L(e)$. We
also assume now that the distribution of velocities in the object
sample is, for a sufficiently large volume, Gaussian with a width
$\sigma_v$:
\begin{equation}
  P(v_\text{r}|\sigma_v) \propto \exp\left(-\frac{v_\text{r}^2}{2 \sigma_v^2}\right)\;.
\end{equation}
 Now we
can build the joint probability of getting $v_\text{r}$,
$\psi_\text{r,rec}$ and $\beta$:
\begin{multline}
  P(v_\text{r},\psi_\text{r,rec},\beta|B,\sigma_v) \\
    \propto P(e(v_\text{r},\psi_\text{r,rec})|B,\sigma_v) P(v_\text{r}|B,\sigma_v)
    P(\psi_\text{r,rec}|B,\sigma_v) \\
    \propto P(\psi_\text{r,rec}|B,\sigma_v) \frac{\exp\left(-\frac{v_\text{r}^2}{2\sigma_v^2}\right)}{1+\left(\frac{v_\text{r} - \beta \psi_\text{r,rec}}{B}\right)^2}\;,
\end{multline}
where the constant of proportionality eventually depends on
$B$, $\sigma_v$ and $\beta$.
Using the theorem of Bayes, it is now possible to compute the
conditional probability that the true velocity of some object is
$v_\text{r}$ given that the reconstructed displacement is
$\psi_\text{r,rec}$:
\vspace{-.1cm}
\begin{multline}
  P(v_\text{r}|\psi_\text{r,rec},\beta,B,\sigma_v) \\ =  \frac{
    \mathrm{e}^{-\frac{v_r^2}{2 \sigma^2_v}} \left(1 +
      \left(\frac{\beta_{*}\psi_r-\alpha_{*} v_r +
          \gamma_{*}}{B_v}\right)^2 \right)^{-1}
    }{
      \int_{v=-\infty}^{+\infty} \mathrm{e}^{-\frac{v^2}{2 \sigma^2_v}} \left(1 +
      \left(\frac{\beta_{*}\psi_r-\alpha_{*} v +
          \gamma_{*}}{B_v}\right)^2 \right)^{-1}\;\text{d}v
    }\;.
\end{multline}
To obtain the total likelihood $\mathfrak{L}(\beta)$ to observe true
velocities $\{ v_{i,\text{r}} \}$ given that the reconstructed
displacements are $\{ \psi_{i,\text{r,rec}} \}$, one may assume the
statistical independence of the $(v_{i,\text{r}},
\psi_{i,\text{r,rec}})$ duets. With this assumption, $\mathfrak{L}$ is
simply
\vspace{-.1cm}
\begin{equation}
  \mathfrak{L}(\beta) = \prod_{i} P(v_{i,\text{r}}|\psi_{i,\text{r,rec},\beta,B,\sigma_v})
\end{equation}
Using that approach we have made measurements in finite volume mock
catalogs. For example, with a PaddedDom reconstruction, one measure
$\Omega_\text{m} = 0.34$ with this approach (for an effective
$\Omega_\text{m}=0.35$ in this catalog), whereas a naive measurement
would yield $\Omega_\text{m} \le 0.26$. 

\vspace{-.3cm}
\section{Conclusion}
\vspace{-.2cm}
We presented a method to predict velocities of galaxies from their
current position.
To solve this problem, we implemented a fast algorithm invented by Dimitri Bertsekas 
\cite{Bertsekas79} and applied the method to a pure dark matter
simulation. It happens that the reconstructed velocities are
impressively accurate on large-scales
(\S~\ref{sec:apply_cosmo}). However, the solution is only approximate 
in regions where multi-streaming occurs. 

We proposed two methods for partially correcting boundary effects (\S~\ref{sec:boundary}) and
showed how boundary effects affect the reconstructed velocity field. 
We preferred the PaddedDom reconstruction scheme as it seems to give
overall better results. Empirically we found that a buffer zone of
20~\Mpch{} is needed before obtaining a reconstructed velocity field
correlated with the one given by the simulation.

At last, we proposed a bayesian model (\S~\ref{sec:stats}) to account
for reconstruction errors while estimating the slope between the
reconstructed displacements and the true velocities of objects in a
galaxy catalogs. 

We would like to continue this work by improving the padding schemes
to have even less boundary effects and make full use of available data
in astronomy. We are also working on an improved algorithm that is
able to take into account in a better way the non-linearities that are
introduced in the velocity field due to gravitational effects occuring
along particle trajectories. This new algorithm will try to fully
solve the Euler-Poisson problem.\footnote{G. Lavaux \& G. Loeper, work
in progress.}

This work is partially supported by the ANR grant BLAN07-2\_183172 (project OTARIE).


\vspace{-.3cm}
\begin{thebibliography}{1}
\expandafter\ifx\csname url\endcsname\relax
  \def\url#1{\texttt{#1}}\fi
\expandafter\ifx\csname urlprefix\endcsname\relax\def\urlprefix{URL }\fi

\bibitem{Peebles89}
P.~J.~E. {Peebles}, {Tracing galaxy orbits back in time}, ApJL 344 (1989)
  L53--L56.

\bibitem{Brenier2002}
Y.~{Brenier}, U.~{Frisch}, M.~{H{\'e}non}, G.~{Loeper}, S.~{Matarrese},
  R.~{Mohayaee}, A.~{Sobolevski{\u i}}, {Reconstruction of the early Universe
  as a convex optimization problem}, MNRAS 346 (2003) 501--524.

\bibitem{Bertsekas79}
D.~P. {Bertsekas}, {A Distributed Algorithm for the Assignment Problem}, MIT
  Press, Cambridge, MA, 1979.

\bibitem{moh2005}
R.~{Mohayaee}, H.~{Mathis}, S.~{Colombi}, J.~{Silk}, {Reconstruction of
  primordial density fields}, MNRAS 365 (2006) 939--959.

\bibitem{CouHydra95}
H.~M.~P. {Couchman}, P.~A. {Thomas}, F.~R. {Pearce}, {Hydra: an Adaptive-Mesh
  Implementation of P 3M-SPH}, ApJ 452 (1995) 797--+.

\bibitem{Lavaux07}
G.~{Lavaux}, R.~{Mohayaee}, S.~{Colombi}, R.~B. {Tully}, F.~{Bernardeau},
  J.~{Silk}, {Observational biases in Lagrangian reconstructions of cosmic
  velocity fields}, ArXiv e-prints 707.

\end{thebibliography}

\end{document}